\documentclass[runningheads]{llncs}
\usepackage{graphicx}
\usepackage[export]{adjustbox}
\usepackage{hyperref}

\usepackage{paralist}

\usepackage{amsmath}
\usepackage{stmaryrd}

\usepackage{cite}

\usepackage{fancyvrb}

\usepackage{booktabs}

\usepackage{multirow}
\newcommand{\nop}[1]{} 

\usepackage{xcolor}

\usepackage[caption=false]{subfig}

\begin{document}
\title{An Empirical Study on the Design and  Evolution of NoSQL Database Schemas}
%

\titlerunning{Design and Evolution of NoSQL Database Schemas}
%
\author{Stefanie Scherzinger \inst{1} \and
Sebastian Sidortschuck \inst{2}}
\authorrunning{Scherzinger and Sidortschuck}
%
\institute{OTH Regensburg, Regensburg, Germany \\
\email{stefanie.scherzinger@oth-regensburg.de} 
\and
OTH Regensburg, Regensburg, Germany\\
\& SPARETECH.io, Stuttgart, Germany\\
\email{sebastian.sidortschuck@sparetech.io}}
\maketitle              
\begin{abstract}
We study  how software engineers design and evolve their domain model when building applications against NoSQL data stores.
Specifically, we target Java projects that use object-NoSQL mappers to interface with schema-free NoSQL data stores. Given the source code of ten real-world database applications, we  extract the {\em implicit}\/ NoSQL database schema.
We capture the sizes of the schemas, and 
investigate whether the schema is denormalized, as is recommended practice in data modeling for NoSQL data stores. Further, we
analyze the entire project
history, and with it, the evolution history of the NoSQL database schema. 
In doing so, we conduct the so far largest empirical study on  NoSQL schema design and evolution.

\end{abstract}
\section{Introduction}

Schema-flexible NoSQL data stores have become popular backends for   building database applications.
Systems like MongoDB  allow for flexible changes to the domain model during application development. In particular,
they have proven themselves in settings where applications are frequently deployed to their production environment, e.g., when web applications are built in an agile approach. 

While the data stores do not enforce a global schema, the application code generally assumes that persisted entities adhere to a certain (if loose) domain model.
Given that schema-flexibility is one of the major selling points of NoSQL data stores,
this raises the question how the domain model, and thereby the implied \emph{NoSQL database schema}, actually evolves. 
We empirically study the dynamics of NoSQL database schema evolution. 
Further, we investigate the question whether the NoSQL database schema is denormalized, as commonly recommended in literature, e.g.~\cite{Sadalage:2012:NDB:2381014}. 

Unfortunately, real-world data dumps of NoSQL data stores are hard to come by. 
We therefore resort to analyzing the source code of applications hosted on GitHub.
We focus on the relevant software stack shown in Figure~\ref{subfig:stack}, namely Java applications that use an object-NoSQL mapper  to store data in either Google Cloud Datastore\footnote{\url{https://cloud.google.com/datastore/}, available since 2009.} or MongoDB\footnote{\url{https://www.mongodb.com/}, available  since 2019}, both popular and mature data stores.
Among over 1.2K open source GitHub repositories with this stack, we have  identified the ten  projects with the largest  NoSQL schemas (a notion introduced shortly). 

Previous studies on schema evolution have primarily focused on schema-full, relational databases~\cite{Curino08schemaevolution,Lin:2009:CEA:1595808.1595817,Qiu:2013:EAC:2491411.2491431,Wu:2011:SEA:2014699.2014942,Skoulis:2015:GUS:2799194.2799242,DBLP:journals/infsof/Sjoberg93}. 
About  NoSQL schema evolution in  real-world applications, little is known that is based on systematic, empirical studies (versus anecdotal evidence): 
Earlier studies  have a different focus (such as the usage of certain mapper features~\cite{Ringlstetter:2016:DME:2896825.2896827}), or analyze a single project (c.f.~\cite{7884653}).

In this paper, we introduce our notion of the NoSQL database schema, which is implicit in object mapper class declarations, even though the underlying NoSQL data stores are schema-free.
In this setting, this paper makes the following contributions:
\begin{compactitem}

\item We formulate three research questions, namely 
(RQ1)~whether the NoSQL database schema is  \emph{denormalized} (as recommended in literature), 
(RQ2) which growth in complexity we can observe in  NoSQL database schemas over the project development time, and  (RQ3)~{\em how}\/ the NoSQL database schema evolves, thereby identifying the common changes.

\item
We analyze the ten projects with the largest NoSQL database schemas among  over 1.2K candidate projects,
based on  static code analysis and the commit history.
We are able to confirm that denormalization is common in NoSQL database schemas. We are further able to show evidence of evolutionary changes to the NoSQL database schema in all analyzed projects. 

\item We discuss our findings w.r.t.\  related studies on relational schema evolution
and find that the {\em churn rate}\/ of NoSQL schemas is comparatively high.
\end{compactitem}

\smallskip
\noindent
\emph{Structure.}
Next, we introduce preliminaries in Section~\ref{sec:prelims}. In Section~\ref{sec:methodology}, we describe our methodology, and state our research questions. In Section~\ref{sec:results}, we present the results of our study, which we then discuss in Section~\ref{sec:discussion}.
We point out threats to the validity of our results in Section~\ref{sec:validity}, and give an overview over related work in Section~\ref{sec:related}. We  conclude with an outlook on future work.

\section{Preliminaries}
\label{sec:prelims}

\begin{figure}[t]
    \centering
    
    \subfloat[The software stack.]{\includegraphics[scale=.315]{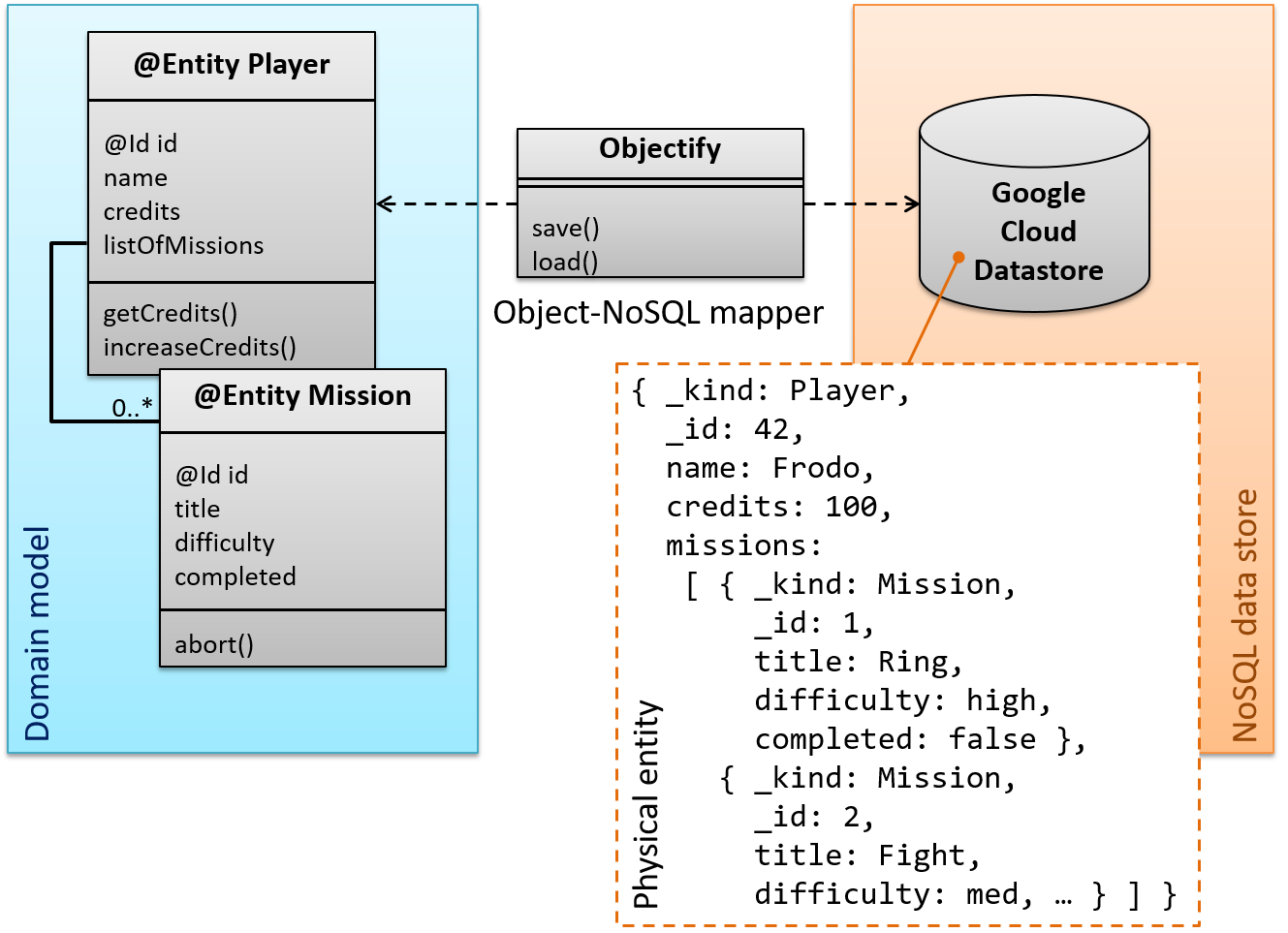}\label{subfig:stack}}%
    \;
    \subfloat[Code changes to class Player.]{\includegraphics[scale=.315]{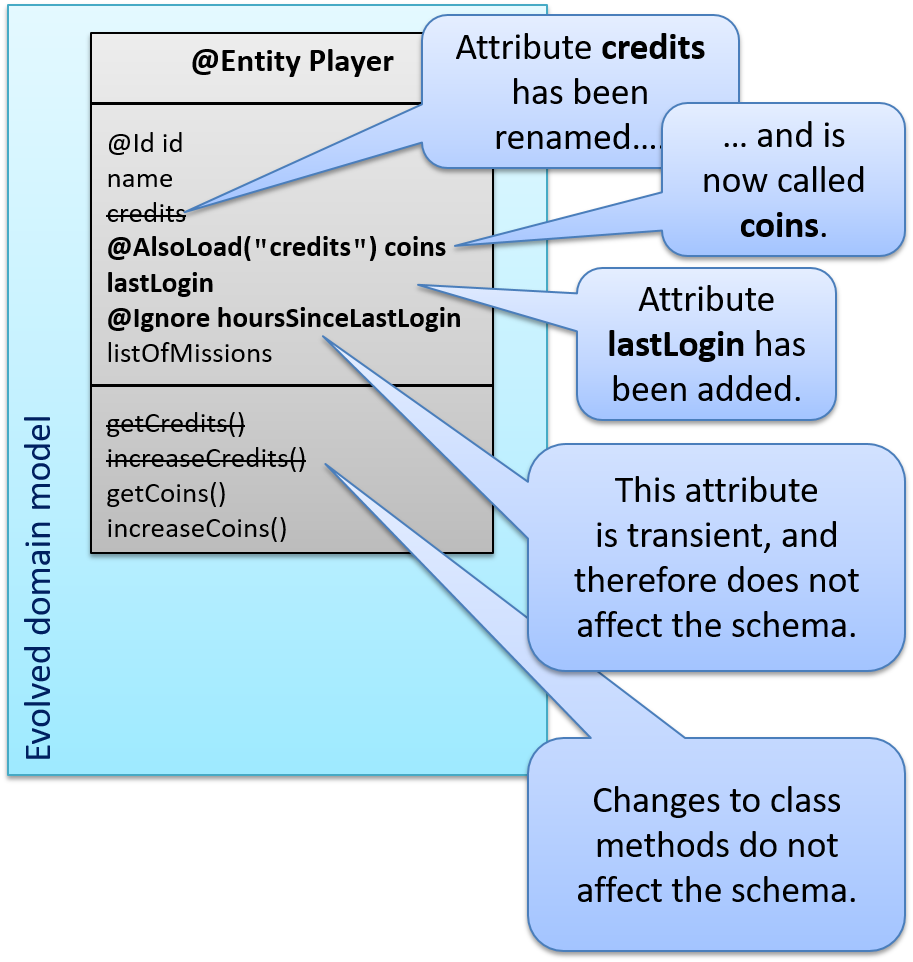}\label{subfig:evolution}}

    \caption{(a) The  object-NoSQL mapper  separates the domain model from the NoSQL data store (adapted from~\cite{Fowler:2002:PEA:579257}). (b) Not all code changes are actually schema-relevant. }
    
    \label{fig:stack}
\end{figure}

We next introduce the software stack studied, as well as our terminology.

\smallskip
\noindent
\emph{Physical entities.}
We consider two popular NoSQL data stores:
 Google Cloud Datastore (called Datastore hereafter) is commercial and hosted on the Google Cloud Platform,
MongoDB is open source. 
Both data stores are schema-free (however, MongoDB offers {\em optional}\/ schema validation). 
Both manage document-like data, which we refer to as the (physical) {\em entities}\/. 
On an abstract level, an entity is a collection of key-value pairs, or {\em properties}\/. Entities may be nested and properties may be multi-valued.
We sketch a Datastore entity representing a player and his or her missions in a  role playing game
 in Figure~\ref{subfig:stack}, in (simplified) JSON notation, to
abstract away from system-proprietary storage formats.
 
\smallskip
\noindent
\emph{Domain models.}
In principle, each  entity in a schema-free data store may have its very own,  unique structure. However, 
in database applications,  
it is safe to assume that the  software engineers have agreed on some {\em domain model}\/, as sketched in Figure~\ref{subfig:stack}. In our setting, the domain model is captured by Java class declarations, yet in the Figure, we use the more compact UML  notation.
(For now, we ignore the \verb!@!-labeled annotations.)
Class  Player declares 
attributes for an identifier, a name, an amount of  credits, and  a list of missions. Each mission also has an identifier, a title, a level of difficulty, and tracks its completion.

\smallskip
\noindent
\emph{Object-NoSQL Mappers.}
Object mappers are state-of-the-art in building data\-base applications~\cite{Fowler:2002:PEA:579257}. Like object-relational mappers,
the object-NoSQL mappers  Objectify\footnote{\url{https://github.com/objectify/objectify}} and Morphia\footnote{\url{https://github.com/MorphiaOrg/morphia}} map Java objects to entities. Objectify is tied to Datastore, and Morphia to MongoDB.
With object-NoSQL mappers, developers merely specify their domain model as Java classes that are annotated with the keyword \verb!@Entity!. Each {\em entity-class}\/ has a unique key (annotated with \verb!@Id!). 
The object mapper provides methods for saving and loading:
In Figure~\ref{subfig:stack},
the class name and the identifying attribute are mapped to the designated properties \verb!_kind! and \verb!_id!.
Objectify maps the player's list of missions to an array of nested entities. %
Yet at application runtime, an entity-class declaration may not  match the structure of  all persisted entities, as discussed next. 

\smallskip
\noindent
\emph{Lazy data migration.}
The data store may also store  legacy versions of  entities.
Figure~\ref{subfig:evolution} shows a new version of entity-class Player, with changes due to new requirements in the software development project. 
Attribute \verb!coins! has replaced \verb!credits!. Merely changing the entity-class in the application code does not affect any existing entities. Instead, persisted entities are only migrated lazily, upon loading:
The new version of  entity-class Player in Figure~\ref{subfig:evolution} is backwards-compatible with  Figure~\ref{subfig:stack}. Once the legacy entity for  Frodo has been loaded, the corresponding Java object will have an attribute \verb!coins!,
as annotation \verb!@AlsoLoad! lazily renames attributes. 

Thus, to obtain a summary of the structural variety of physical entities in the data store (based on code analysis alone, not having access to the data store contents itself), we need to consider the entire evolution history of entity-classes.

\smallskip
\noindent
\emph{NoSQL database schema evolution.}
We base our notion of the {\em NoSQL database schema}\/ (or shorter, \emph{NoSQL schema}) on the domain model. 
This idea of treating entity-classes as schema declarations is re-current in literature, c.f.~\cite{DBLP:conf/modelsward/SevillaMM17,DBLP:conf/icde/ScherzingerCA15}.
Note that not all Java attributes are relevant for the NoSQL database schema: Attributes that are transient, e.g., carrying Objectify annotation \verb!@Ignore!, are not schema-relevant: 
%
%
The value of \verb!hoursSinceLastLogin! 
in Figure~\ref{subfig:evolution}
is not persisted (it may be derived from  \verb!lastLogin!).
Also, class methods are not schema-relevant. Thus, code changes that only affect transient attributes or class methods are part of software evolution, but not of schema evolution. Therefore, they are not  considered {\em schema changes}\/ by us.

\smallskip
\noindent
\emph{Denormalized entity-classes.}
The recommendation in working with  Datastore and MongoDB is to intentionally denormalize the schema.\footnote{E.g.\  ``6 Rules of Thumb for MongoDB Schema Design'' at  \url{https://www.mongodb.com/blog/post/6-rules-of-thumb-for-mongodb-schema-design-part-2}, June 2015.}
This can be done by either nesting entities, or by using multi-valued properties, such as the array of \verb!Mission!s in Figure~\ref{fig:chage:dist}.
There are various motivations for denormalization, one being that traditionally, the query languages do not provide a join operator (such is still the case in Google Cloud Datastore, and this also used to be the case for MongoDB), so joined data is materialized in the data store. Another reason is that transactions between an arbitrary number of entities may not be  supported\footnote{We point to the concepts such of entity groups and cross-group transactions in the \emph{classic} Google Cloud Datastore~\cite{Sanderson:2015:PGA:2846415}, which is in the process of being deprecated.}. Consequently, transactionally safe updates are often realized by updates to a single, aggregate entity. 

In the following, we say an entity-class is denormalized if it does not declare flat, relational-style tuples in first normal form, i.e., with atomic attribute values only.
So unless all schema-relevant attributes have Java primitive types (such as  \verb!Integer!, \verb!String!,  \verb!Boolean!, \dots ), we say the entity-class is denormalized.
As we discuss in Section~\ref{sec:analysis_process}, this is a practical yet  conservative approach.

As an example, the entity-class declarations for players, sketched in Figure~\ref{fig:stack}, is denormalized, due to the multi-valued attribute \verb!listOfMissions!.

\section{Methodology}
\label{sec:methodology}

In the following, we describe our methodology, such as the context of our analysis, the research questions, and the analysis process.
While our outline has strong analogies to
Qiu et al.~\cite{Qiu:2013:EAC:2491411.2491431} and their analysis of  {\em relational}\/ schema evolution, our   process is rather different: we cannot analyze schemas declared in a declarative data definition language, such as SQL. Rather, we need to parse raw Java code.

\subsection{Context}
\label{sec:context}

We used BigQuery%
\footnote{Google BigQuery is a commercial cloud service. This data warehousing tool allows for querying the GitHub open data collection, mostly non-forked projects with an open source license: \url{https://cloud.google.com/bigquery/}.}
to identify relevant open source repositories on GitHub,
as of September 4th, 2018. We consider a  repository (which we synonymously refer to as a project) relevant if it contains Java import statements for Objectify or Morphia. 
We cloned over 1.2K candidate repositories and excluded any repositories that
(1)~have fewer than 20 commits (to exclude tinker projects),
(2)~are the Morphia or Objectify source code (or forks thereof), 
(3)~or are flagged as forks from  repositories already covered, with no schema-relevant code changes after the fork. 
We analyze the project history using \verb!git log!\footnote{We state the exact command pattern for reproducability: {\tt \scriptsize git log --before=2018-09-04T00:00:00 --cherry-pick --date-order --pretty=format:"\%H;\%aI;\%cI;\%P"}.}. This allows us to re-trace the development history of all entity classes. We parse and aggregate the log output using Python scripts.

Among all projects analyzed, we determined the maximum number of entity-classes throughout the project history, and settled on the top-5 projects for Objectify and Morphia respectively.
Table~\ref{table:project_info} lists these projects with their life cycles up to the latest commit at the time of our analysis. 
We also state the total number of commits at the time.
We state the minimum and maximum number of entity-classes throughout the project history, as well as the total number of lines of code between the first and last analyzed commit (measured with \verb!cloc!\footnote{\url{https://github.com/AlDanial/cloc}} and reported in thousands).

\begin{table}[t]
\scriptsize
\caption{Characteristics of the studied database applications.}
\setlength{\tabcolsep}{0.2em}
\begin{tabular}[]{l|lrrrr}
\toprule
 \multicolumn{1}{r}{} & \textbf{Project} & \textbf{Life Cycle} & \textbf{\# Commits} & \begin{tabular}[c]{@{}c@{}}\textbf{\# Entity-}\\\textbf{classes}\end{tabular} & \textbf{LoC (K)}\tabularnewline
\midrule
\parbox[t]{4mm}{\multirow{4}{*}{\rotatebox[origin=c]{90}{\textbf{Objectify}}}}& Cryptonomica/cryptonomica &  04/16 $\sim$ 09/18 & 185 & 0 $\sim$ 29 & 0 $\sim$ 526\tabularnewline 
& FraunhoferCESE/madcap &  12/14 $\sim$ 03/18 & 853 & 0 $\sim$ 82 & 0 $\sim$ 17\tabularnewline 
& google/nomulus &  03/16 $\sim$ 09/18 & 2,025 & 51 $\sim$ 55 & 138 $\sim$ 224\tabularnewline 
& nareshPokhriyal86/testing &  01/15 $\sim$ 02/15 & 25 & 0 $\sim$ 79 & 0 $\sim$ 449\tabularnewline 
& Nekorp/Tikal-Technology &  04/15 $\sim$ 11/15 & 59 & 0 $\sim$ 43 & 0 $\sim$ 49\tabularnewline 
\midrule
\parbox[t]{4mm}{\multirow{4}{*}{\rotatebox[origin=c]{90}{\textbf{Morphia}}}}& altiplanogao/tallyframework &  06/15 $\sim$ 06/16 & 167 & 0 $\sim$ 24 & 0 $\sim$ 5\tabularnewline 
& bujilvxing/QinShihuang &  10/16 $\sim$ 12/16 & 154 & 0 $\sim$ 36 & 0 $\sim$ 21\tabularnewline 
& catedrasaes-umu/NoSQLDataEngineering &  11/16 $\sim$ 09/18 & 711 & 0 $\sim$ 28 & 0 $\sim$ 280\tabularnewline 
& GBPeters/PubInt &  10/16 $\sim$ 02/18 & 69 & 0 $\sim$ 27 & 0 $\sim$ 5\tabularnewline 
& MKLab-ITI/simmo &  07/14 $\sim$ 02/17 & 142 & 0 $\sim$ 51 & 0 $\sim$ 5\tabularnewline 
\bottomrule
\end{tabular}
\label{table:project_info}
\end{table}

\subsection{Research Questions}
\label{sec:research_questions}

%
\begin{compactdesc}
\item [RQ1: Are NoSQL schemas denormalized?]

We analyze the structure of entity-classes, whether they map to flat tuples in first normal form, 
or whether they represent denormalized data.

\item [RQ2: What is the growth in complexity of the NoSQL schema?] 

We capture schema complexity
based on metrics recognized in literature.

\item [RQ3: How does the NoSQL schema evolve?]

We automatically identify and classify evolutionary changes to the NoSQL schema.

\end{compactdesc}



\subsection{Analysis Process}
\label{sec:analysis_process}

\smallskip
\noindent
\emph{Locating entity-classes.}
We replay the commit histories and
use the Java parser QDox\footnote{\url{https://github.com/paul-hammant/qdox}} to parse class declarations.
We identify entity-classes by the object mapper annotation \verb!@Entity!, which may also be inherited.%
\footnote{In earlier versions of the mapper libraries, this annotation was only optional, so it cannot be relied upon. We therefore also search for the mandatory annotation~{\tt @Id}, and thus reliably detect polymorphic entity-classes. (c.f.\ Section~\ref{sec:validity}).}

\smallskip
\noindent
\emph{Denormalization.}
To determine whether an entity-class is denormalized, we parse its Java declaration and strip away attributes that are not relevant to the NoSQL schema. We then analyze the types of the remaining attributes. Unless all have primitive types (such as \verb!Integer!, \verb!String!, or \verb!Boolean!), we assume that the entity-class is denormalized. 

In most cases, we correctly recognize denormalization:
(1)~if the entity class declaration contains container classes   (e.g., a Java \verb!Collection!), and therefore an attribute is multi-valued. (2)~Equally, the entity-class may contain nested entity classes, giving it a hierarchical structure. 

However, there are also cases where this approach is a conservative simplification, and we might falsely categorize an entity-class as denormalized: 
an attribute type may be declared in a  third-party library, which is inaccessible to us (see also our discussion in Section~\ref{sec:validity}). Also, an attribute type may be a custom type that the developers declared. To realize that a custom type is just a wrapper for a basic Java type, we would have to run more involved code analysis. Yet typically, polymorphic types are involved, and we are confronted with the inherent limitations of static code analysis.

\smallskip
\noindent
\emph{Identifying schema changes.}
We identify commits with schema-relevant changes by comparing succeeding versions of the application source code:
We register when (1)~a new entity-class is added
or an  entity-class is removed,
(2)~a schema-relevant attribute is added or removed in an entity-class declaration,
 and (3)~further, changes to schema-relevant attributes, such as to their types, default initializations, or even object mapper annotations.
We  only focus on   changes which we can recognize programmatically.
Recognizing renaming or splitting an entity-class, or renaming an attribute, are instances of the challenge of schema matching and mapping~\cite{Bellahsene:2011:SMM:1972526},
and cannot be fully automated.

\section{Results of the Study}
\label{sec:results}


\begin{figure}[tbp]
    \centering

    \subfloat[Objectify-based projects.]{\includegraphics[width=\columnwidth]{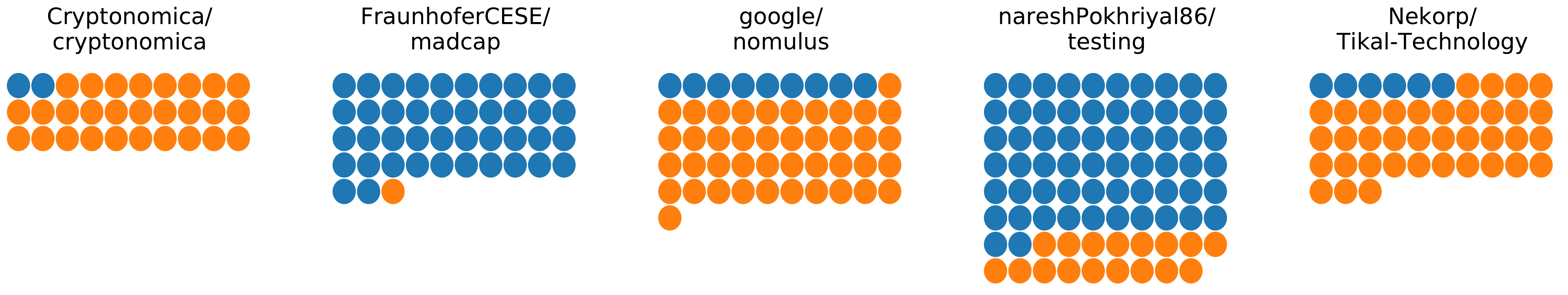}\label{sub:nf:objectify}}%
    \;
    \subfloat[Morphia-based projects.]{\includegraphics[width=\columnwidth]{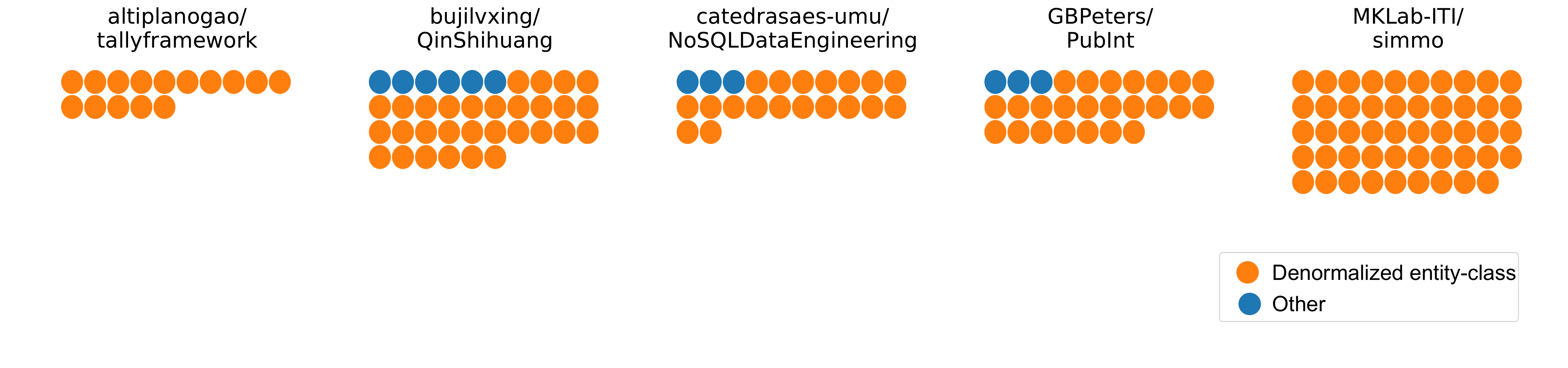}\label{sub:nf:morphia}}

    \caption{Visualization of denormalized NoSQL database schemas.}
    \label{fig:normalforms}
\end{figure}

\subsection{RQ1: Are NoSQL schemas denormalized?}

We analyze the entity-class declarations in their most current version w.r.t.\ denormalization. 
The results are visualized in 
Figure~\ref{fig:normalforms}.
For each analyzed project, we 
show a dot matrix chart. The number of dots represents the number of entity-classes.  The brighter (orange) dots represent the  entity-class declarations which we must assume to be denormalized, due to the limits of static code analysis. 
The darker (blue) dots represent the other entity-classes.

Notably, each project contains at least one denormalized entity-class, so all schemas are denormalized.
With the exception of two Objectify-based projects,  denormalized entity-classes dominate the NoSQL database schemas. 
There are even two Morphia-based projects where all entity-classes are denormalized.

\smallskip
\noindent
\emph{Results.} We find that each project analyzed has denormalized entity-classes in its NoSQL schema. This shows that developers make active use of denormalization. However, without qualitative studies based on developer surveys, we do not know whether (1) the developers   
 consciously chose a database which allows for a denormalized database schema, as this better suits their conceptual model. However, it could also be that (2)~they are actually forced denormalize their data  model, due to 
 the technological limitations of NoSQL data stores 
 (briefly discussed in Section~\ref{sec:prelims}).


\begin{figure}[tb]
    \centering

    \subfloat[Objectify-based projects.]{\includegraphics[width=\textwidth]{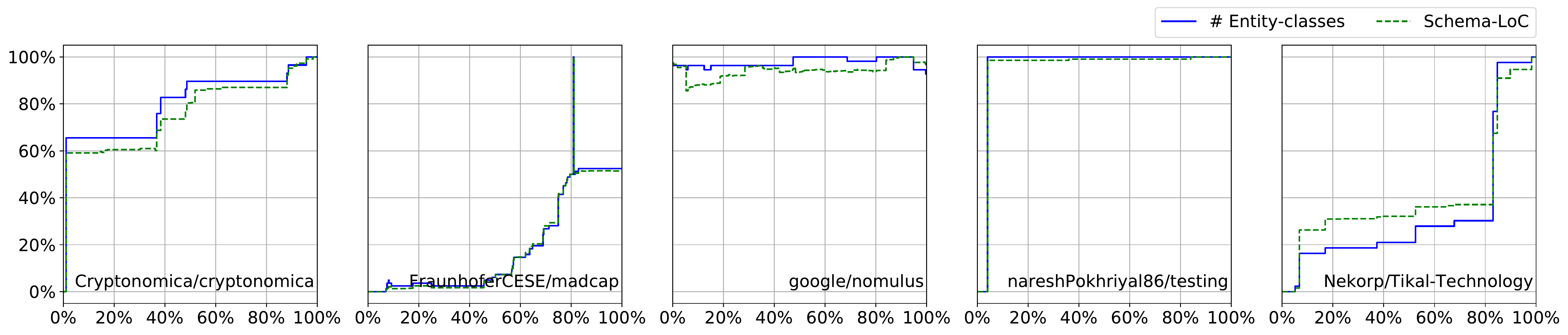}\label{subfig:evol:objectify}}%
    \;
    \subfloat[Morphia-based projects.]{\includegraphics[width=1.01\textwidth]{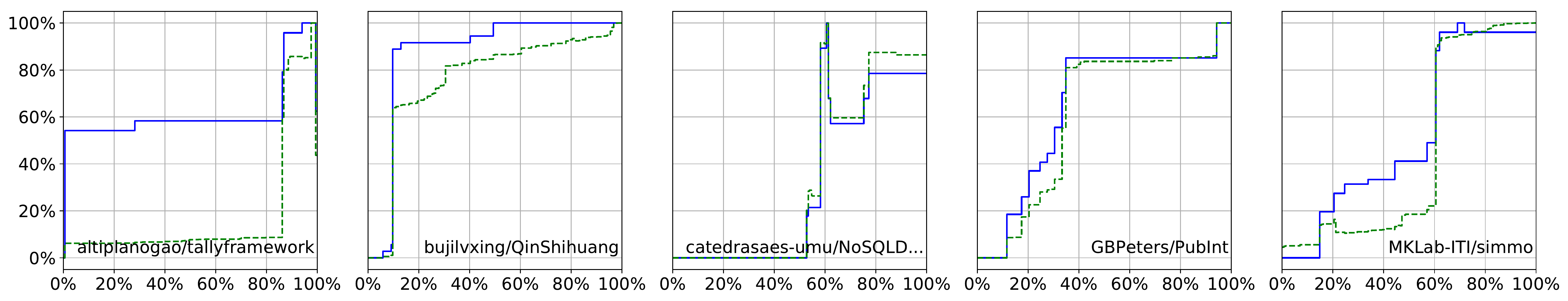}\label{subfig:evol:morphia}}

    \caption{Evolution trend of entity classes. The horizontal axes show the project progress, in percentage of commits analyzed. The vertical axes show the complexity of the schema w.r.t.\ its maximum, for two alternative metrics. (Visualization modeled after~\cite{Qiu:2013:EAC:2491411.2491431}.)}
    \label{fig:evolution_trend}
\end{figure}

\subsection{RQ2: What is the growth in complexity  of the NoSQL schema?}

In empirical studies on relational schema evolution, the number of tables is considered a simple approximation for schema complexity~\cite{DBLP:conf/icde/JainMH16}. Accordingly, we track the number of entity-classes over time in 
Figure~\ref{fig:evolution_trend} (based on a visualization idea from~\cite{Qiu:2013:EAC:2491411.2491431}).
For each project, one chart is shown. On the horizontal axis, we track the progress of the project, measured as the percentage of \verb!git! commits analyzed. For the \verb!madcap! project, this is based on 853 commits (c.f.\ Table~\ref{table:project_info}).
On the vertical axis, we track the size of the NoSQL database schema using two metrics. One is the number of entity classes (blue solid line). This metric is also normalized w.r.t.\ its maximum throughout the project history.
So for \verb!madcap!, the 100\% peak corresponds to 82 entity classes, some of which were removed in the later phase of the project. 

The second line denotes a  ``proxy metric''~\cite{DBLP:conf/icde/JainMH16} for approximating the size of the NoSQL schema, where we count the lines of code of the entity-classes (including superclasses,
excluding comments and empty lines), and thereby compute the {\em Schema-LoC}\/.%
\footnote{We find this proxy-metric preferable over counting (schema-relevant) attributes, as is common in studies on relational schema evolution: (1)~Entity-classes with more schema-relevant 
attributes have more lines of code accordingly. 
(2)~In static code analysis, we cannot reliably count nested attributes: Abstract container classes and the use of polymorphism in general, make it impossible to know the number and nature of nested attributes at compile time. With Schema-LoC, we are able to abstract from this issue.}
There is shrinkage,
yet overall, schema complexity increases.

\smallskip
\noindent
\emph{ Results.}
{\bf 1)}~As in the study by Qiu et al.\ on relational software evolution~\cite{Qiu:2013:EAC:2491411.2491431}, we can confirm that while the projects differ in their life-spans and commit activity, in nearly all projects, the NoSQL schema grows over time. However, there may be phases of refactoring, leading to dips in the curves. 
{\bf 2)}~Apparently, Schema-LoC lends itself nicely as a proxy-metric, and we obtain high correlation coefficients when comparing to the number of entity classes.
%
%
As Schema-LoC depends on the number of attributes in an entity class,
we can retrace an effect reported in~\cite{Qiu:2013:EAC:2491411.2491431}, namely that  entity-classes and their attributes (corresponding to tables and columns) have largely analogous dynamics.
{\bf 4)}~In general, the schema grows more than it shrinks.
This is in line with studies on relational schema evolution.
{\bf 5)}~One observation in~\cite{Qiu:2013:EAC:2491411.2491431} was that the schema stabilizes early: There, for 7 out of 10 projects, 60\% of the maximum number of tables is reached in the first 20\% of the commits. 
Interestingly, in our study, the number of entity-classes reaches the 60\% in only 4 projects.
{\bf 6)}~In~\cite{Qiu:2013:EAC:2491411.2491431}, less than 2\% of all commits contain valid schema changes (across all ten projects analyzed there). In our study, the share of commits with schema-relevant changes 
is between 2.8\% and over 30\%, with 4 projects reaching over 20\%. Clearly, we observe higher churn rates.

\nop{
altiplanogao/tallyframework          | 20 | 11.976047904191617 %

bujilvxing/QinShihuang               | 48 | 31.16883116883117 %

catedrasaes-umu/NoSQLDataEngineering | 22 | 3.0942334739803097 %

GBPeters/PubInt                      | 14 | 20.28985507246377 %

MKLab-ITI/simmo                      | 44 | 30.985915492957744 %

Cryptonomica/cryptonomica            | 22 | 11.891891891891893 %

FraunhoferCESE/madcap                | 48 | 5.627198124267292 %

google/nomulus                       | 57 | 2.814814814814815 %

nareshPokhriyal86/testing            | 2  | 8.0 %

Nekorp/Tikal-Technology              | 12 | 20.33898305084746 %
} 

\begin{figure}[t]
    \centering
    \includegraphics[width=1\textwidth]{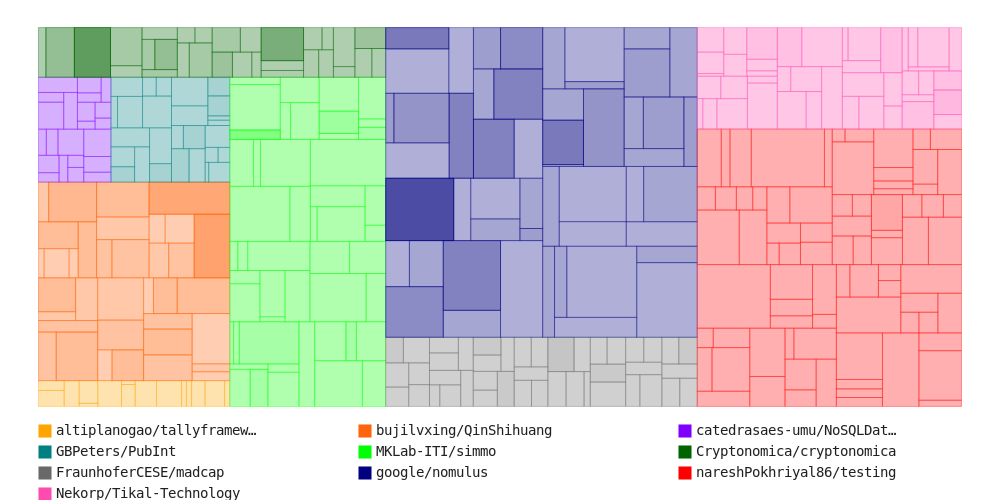}
    \caption{Visualizing relative schema sizes and churn. Each rectangle represents an entity-class, its area proportional to its size in lines of code (specifically Schema-LoC). The hue represents the relative frequency of schema changes within the same project.}
    \label{fig:treemap}
\end{figure}

\begin{figure}[t!]
    \centering
    
        \subfloat[By project.]{\includegraphics[width=\textwidth]{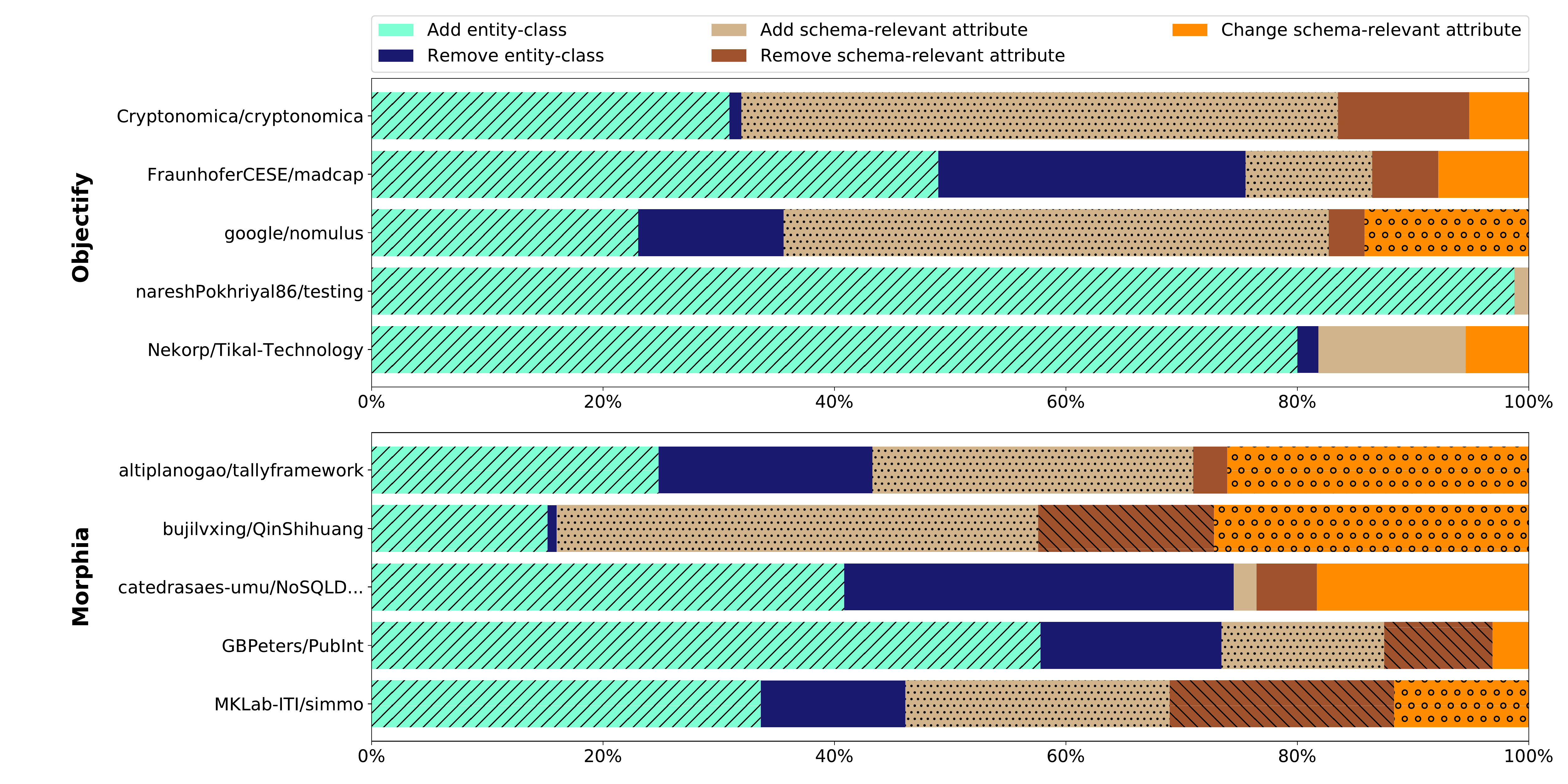}\label{subfig:dist:chart}}%
    \;
    \subfloat[Objectify-based vs.\ Morphia-based projects.]{
    \scriptsize
    \begin{tabular}{l r r  r  r r r}
    \toprule 
   & \multicolumn{2}{c}{\bf Entity-class } & &
  \multicolumn{3}{r}{\bf Schema-relevant attribute}\\
    & \bf \quad  add  & \bf  remove & & \bf add  & \bf remove & \bf change \\

    \midrule
   \bf Objectify & 56.3\% & 8.4\% & \quad \quad \quad & 24.7\% & 4.0\% & 6.5\%\\
   \bf Morphia & 34.5\% & 16.2\% & & 21.6\% & 10.4\% & 17.2\% \\
    \midrule
    \bf Overall & 34.8\% & 15.1\% & & 27.3\% & 7.4\% & 15.4 \% \\
    \bottomrule
    \end{tabular}
    \label{subfig:dist:table}
   } 

\subfloat[Drill-down into the remaining changes to schema-relevant attributes.]{
\scriptsize
\begin{tabular}[]{l|lrrrrr}
\toprule
\multicolumn{1}{r}{} & \textbf{Project} & \textbf{Type} & \textbf{Initialization} & \textbf{Annotations} \tabularnewline
\midrule
\parbox[t]{4mm}{\multirow{4}{*}{\rotatebox[origin=c]{90}{\textbf{Objectify}}}}& Cryptonomica/cryptonomica &  2 & 0 & 3\\ 
& FraunhoferCESE/madcap &  9 & 0 & 6\\ 
& google/nomulus &  11 & 2 & 58\\ 
& nareshPokhriyal86/testing &  0 & 0 & 0\\ 
& Nekorp/Tikal-Technology &  0 & 0 & 3\\ 
\midrule
\parbox[t]{4mm}{\multirow{4}{*}{\rotatebox[origin=c]{90}{\textbf{Morphia}}}}& altiplanogao/tallyframework &  7 & 3 & 54\\ 
& bujilvxing/QinShihuang &  32 & 33 & 15\\ 
& catedrasaes-umu/NoSQLDataEngineering &  43 & 0 & 13\\ 
& GBPeters/PubInt &  0 & 2 & 2\\ 
& MKLab-ITI/simmo &  7 & 5 & 18\\ 
\bottomrule
\end{tabular}
\label{table:property_changes}
}

 \caption{Distinguishing different kinds of schema changes: (a) and (b): Relative shares of schema changes (by project and by mapper library). (c) Zooming in on the remaining changes in schema-relevant attributes mentioned in (a), showing absolute values.}
 \label{fig:chage:dist} 
 \end{figure}

\subsection{RQ3: How does the NoSQL schema evolve?}

We first investigate how often  entity-classes undergo schema changes when compared to others inside the same project, and how large they are in terms of our proxy metric Schema-LoC. 
Figure~\ref{fig:treemap} visualizes the entity classes making up the ten NoSQL schemas as a tree map. This figure is best viewed in color. Each colored area represents one project. Inside, each rectangle represents one entity-class, the area proportional to its Schema-LoC. Darker hue indicates that an entity-class has undergone more schema changes than the other entity-classes in the same project. For instance, for \verb!nomulus!, the darkest area represents~12 schema changes against the same entity-class. Thus, some entity-classes change quite more often than others. However, there are also projects where schema changes affect entity-classes quite uniformly.

In Figure~\ref{fig:chage:dist}, we capture the distribution of schema changes according to the kind of change. 
In Subfigure~\ref{subfig:dist:chart}
 (after Qiu et al.\ in~\cite{Qiu:2013:EAC:2491411.2491431}), we break down the distribution of changes by project.
Note that when a new entity-class is added, we do {\em not}\/ count this as adding attributes at the same time. 
Notably, the distributions are project-specific.
We now discuss two projects that stand out.

In the fourth Objectify-based project, adding an entity-class makes up for nearly all changes. Considering the project characteristics in Table~\ref{table:project_info} reveals that this project is an outlier in several regards: With only 25 commits, it has barely made the bar for being considered in our analysis (see Section~\ref{sec:context}).
At the same time, this project holds the second largest number of entity-classes in any project considered in this analysis. Since the life cycle considered is only two months, this project is in a very early stage of development at the time of this analysis. Thus, it seems plausible that at this early phase, the developers kick start their data model by declaring the entity classes in bulk. 

In contrast, the second Morphia-based project stands out as the project with the least share of entity class additions. Since the \verb!git! commit messages are in Chinese (which the authors of this paper do not master), we find it difficult to retrace the developers' motivation. What is noticeable in Subfigure~\ref{subfig:evol:morphia} is that while the number of entity classes increases in less than 10 distinct steps, the proxy metric Schema-LoC changes in more fine-granular steps. Thus, the entity-classes undergo more frequent changes. This matches the distribution plotted, as then the share of entity-class creations is smaller by comparison.
Subfigure~\ref{subfig:dist:table} summarizes Subfigure~\ref{subfig:dist:chart}, and aggregates the changes by the mapper library. While we see  project-specific fluctuations, when we group by mapper library, we also observe differences in the distribution. Overall, additions (whether of entity classes or of schema-relevant attributes) dominate.

In Table~\ref{table:property_changes}, we break down the schema-relevant attribute changes  listed in Subfigures~\ref{subfig:dist:chart} and~\ref{subfig:dist:table}:
{\bf (a)}~For some projects, types change. {\bf (b)}~For others, the  initialization changes.
A drill-down reveals that (as may be expected) adding an initial value is the most frequent change, followed by changing the initialization value.
{\bf (c)}~In other cases, mapper annotations that affect the schema are added or removed.
The most frequent annotations added are \verb!@PersistField! and
\verb!@Reference!. The first is from a third-party framework. Since it is schema-relevant, we report it. 
The second supports referential constraints.
Sporadically, third-party annotations are added to declare additional  constraints, such as \verb!@Min!. 





\smallskip
\noindent
\emph{ Results.}
{\bf 1)}~We can confirm the observations from related work on relational schema evolution that schema changes are generally not  distributed uniformly~\cite{Qiu:2013:EAC:2491411.2491431,journals/is/VassiliadisZS17}. 
{\bf 2)}~As already observed for RQ2, the trend is that entity-classes are added more frequently than they are removed. We see a similar pattern for schema-relevant attributes,  in line with studies on relational schema evolution.
Overall, in 9 out of 10
projects, additions collectively account for more than 50\% of the changes. 
In~5 projects, they even account for over 70\% of the changes. 
{\bf 3)}~While additions are generally more frequent, there are also projects where removals of entity classes occur to a non-significant degree. Related work on relational schema evolution has shown that there are what the authors call survivor tables~\cite{DBLP:conf/caise/VassiliadisZ17}, whereas there are that are more short-lived. The observation that entity-class removals are very project specific has also been made in~\cite{Qiu:2013:EAC:2491411.2491431}.
{\bf 4)}~Among all annotation changes,
only 15 concern referential constraints (annotation \verb!@Reference!). The authors of two relate studies on relational schema evolution, both~\cite{Qiu:2013:EAC:2491411.2491431} and~\cite{DBLP:journals/computing/VassiliadisKZZ19}, have observed that changes concerning referential integrity constraints are also rare in relational schema evolution. With  NoSQL data stores, this is  to be expected, as referential integrity is not supported to the same extent. 
%
{\bf 4)}~While Qiu et al.~\cite{Qiu:2013:EAC:2491411.2491431} found changes in attribute types to be the number one change for half of the projects analyzed (even outnumbering additions of either tables or columns), we do not see evidence of this effect here. 


\section{Discussion}
\label{sec:discussion}

We can reproduce the main results  from related work on relational schema evolution: There is strong evidence of NoSQL schema evolution, and additions are dominant schema changes.
However,
we do not see the schema stabilizing in the early phases of all projects, which may partly be due to shorter project life spans:
The ten projects studied in~\cite{Qiu:2013:EAC:2491411.2491431} are PHP applications backed by relational databases, and have longer life cycles (two with ten years), more commits (starting at nearly 5K), and more lines of code. This is to be expected with a much older and thus more widely adopted stack.

Still, we  do suspect 
that NoSQL developers  evolve their schema more continuously. One indicator supporting this hypothesis is that we see higher {\em churn rates}\/, so a larger share of the commits contains code changes that affect the schema.
This calls for further study.
Due to this churn, making sure that entity-class declarations are ``backwards'' compatible with legacy entities, persisted by earlier versions of the application code, may become an overwhelming task.
There are first proposals for assisting tools, e.g., by type-checking versions of entity-class declarations~\cite{DBLP:conf/icse/CerqueusAS15}. Clearly, more research is needed on systematic tool support.

The fact that denormalization is common shows that solutions for managing relational schema evolution, managing flat tuples,  will not transfer immediately. Rather, when devising frameworks, we may want to turn to related work on frameworks for handling schema evolution in XML (e.g.~\cite{DBLP:journals/informaticaLT/KlimekMNH15}) or object-oriented databases (e.g.~\cite{796507}) for inspiration on what has shown to be feasible.

\section{Threats to Validity} \label{sec:validity}

\noindent
\emph{Construct validity.}
%
{\bf (1)}~With applications using older versions of Objectify and Morphia, we cannot rely on the \verb!@Entity!-annotation to identify entity-classes, so we also consider the \verb!@Id! annotation. To be confident that this does not lead to false positives, we performed manual checks. (With Objectify, we cross-checked which entity-classes were registered with  \verb!ObjectifyService!, a mandatory programming step.) 
{\bf (2)}~In static analysis, we encounter a limitation with attribute types from third-party libraries. 
Tracking down these libraries is out of scope (and not even possible in all cases). 
Thus, there are attributes that are not fully captured by Schema-LoC.
Yet as this is a proxy-metric to start with, we consider this threat acceptable.
Third-party libraries also affect the recognition of entity-classes as denormalized. Having sampled and inspected the entity-class declarations, we are confident that -- given the limitations of static code analysis -- the risk of false positives is acceptable.
{\bf (3)}~We treat each single commit as contributing to a new version of the schema. 
There are software development teams that operate by continuous deployment, so tested code is immediately and autonomously deployed to the production environment. There, in theory, each commit containing a schema change comprises  a new schema version.
Yet rather often, a release to production comprises more than one commit. Unfortunately, we are not able to tell in static code analysis which commits where released when. There are development teams that tag  release commits, but this is project-internal culture, and not consistently the practice across all ten studied projects. Therefore, we must go by the simplifying assumption that each commit  declares a new NoSQL database schema.

\smallskip
\noindent
\emph{External validity.}
We next discuss threats in generalizing our results to other software stacks.
{\bf (1)}~It would be desirable to search additional code repositories, and extend to further NoSQL data stores, object mapper libraries, and programming languages.
{\bf (2)}~Extending our analysis to projects that do not use object-mappers requires a different kind of static code analysis, and was implemented in a related study that involved a single MongoDB project~\cite{7884653}. 
At the same time, object mappers are state-of-the art in modern application development, and by now, Objectify and Morphia are actually part of official Datastore and MongoDB tutorials (even though they started as independent projects).
Thus, we do analyze a highly relevant stack.
{\bf (3)}~There is the fundamental question whether studies on open source projects generalize to commercial projects. 

\section{Related Work}
\label{sec:related}

Database schema evolution is a timeless research area, with various proposals how to systematically manage schema changes. Providing tool support, however, is not the scope of this paper. In the following discussion of related work, we therefore focus on empirical studies on schema evolution in open source projects.

It is only natural that
the  availability of public code repositories has  enabled empirical studies on relational schema evolution~\cite{Wu:2011:SEA:2014699.2014942,Curino08schemaevolution,Lin:2009:CEA:1595808.1595817,Qiu:2013:EAC:2491411.2491431,Skoulis:2015:GUS:2799194.2799242,DBLP:journals/infsof/Sjoberg93,DBLP:conf/caise/SkoulisVZ14,DBLP:conf/er/VassiliadisZS15,DBLP:conf/caise/VassiliadisZ17}.
Among their key findings, these studies show that the schema evolves. They confirm that adding tables or columns are frequent changes. 
In these settings,  the schema is specified declaratively (usually in SQL). Accordingly, the term  {\em schema modification operations} (SMOs)~\cite{Curino08schemaevolution} does not transfer well to our stack. Rather than declarative DDL statements, we need to parse raw  Java code:
While the authors of~\cite{Wu:2011:SEA:2014699.2014942} also parse application code, they do so to extract declarative statements embedded in code.

So far, there are only few empirical studies on schema evolution in NoSQL data stores.
Our work builds on an earlier analysis~\cite{Ringlstetter:2016:DME:2896825.2896827} on the adoption of mapper annotations for lazy schema evolution, which is a different focus.
The authors in~\cite{7884653} present an approach for identifying a schema evolution history in MongoDB-based Java applications. Different from us, the authors do not assume that an object-NoSQL mapper is used to access the data store. Rather, they analyze direct calls to the MongoDB API. The schema derived is similar to our notion of the NoSQL schema, since it captures the perspective of the application code. The authors evaluated their approach for a single open source project, whereas our study has a broader basis, considering ten projects. Moreover, their contribution is to derive a visualization of the schema evolution history.

Meanwhile, there is a growing body of work on extracting schema descriptions~\cite{Baazizi2019, DBLP:conf/btw/KlettkeSS15,DBLP:conf/modelsward/SevillaMM17} 
from large collections of JSON data. While this a bottom-up approach, starting from the data, we proceed top-down, analyzing application code.

In capturing schema complexity based on Java class declarations, we could have resorted to software metrics~\cite{Lanza:2010:OMP:1965070}. However, it is not clear how metrics indicating an overly complex object-oriented design (e.g.\ classes with many attributes) transfer. The practice of building aggregate models in NoSQL schema design may actually be orthogonal.

We refer to~\cite{DBLP:conf/icde/JainMH16} for a high-level discussion on schema variety versus code variety, as well as metrics for programmatic schema analysis.


\section{Conclusion and Outlook}
\label{sec:conclusion}

In this paper, we present the study on NoSQL schema evolution with the largest data basis so far, analyzing ten real-world, open source projects.  
We track the schema growth as well as the nature of changes to the NoSQL schema. 
We are able to reproduce most of the insights of related studies on relational schema evolution, but we have also identified subtle differences.

Since this is a first systematic study,  many interesting questions remain unanswered. We remark on two.
(1)~Originally, we set out to compile detailed statistics on the structure of denormalized  entity-classes, such as their nesting depth. However, we found that Java code written by experienced developers (e.g., as is the case with Google's \verb!nomulus! project) is highly polymorphic. This makes it impossible to compute reliable statistics based on the static analysis of entity-class declarations. However, more holistic analysis techniques, such as data flow analysis of the entire application code, might reveal further insights.
(2)~ We see evidence that the schema evolves, but we do not know the factors that influence NoSQL schema evolution. This calls for follow-up work, where we take the \verb!git! commit messages into account, which often comment the reason for a schema change. What is also needed are qualitative studies, surveying developers who routinely deal with NoSQL schema evolution.















\paragraph*{Acknowledgements}
%
This project was funded by the {\em Deutsche Forschungsgemeinschaft}\/ (DFG, German Research Foundation), grant number \#385808805. 
%
%

\enlargethispage{2cm}
%
%
\bibliographystyle{splncs04} 

\bibliography{bib}

\end{document}